# Electric Transport Properties in the 2D-MoS$_2$

V.V. Vainberg, O.S. Pylypchuk[*], V.N. Poroshin, M.V. Olenchuk, G.I. Dovbeshko

Institute of Physics, NAS of Ukraine, 46 Nauki Prosp., 03068 Kyiv, Ukraine

[*]*pylypchuk@iop.kiev.ua*

**Abstract**

The features of electric transport in the powder samples consisting of the nanosize 2D-MoS$_2$ flakes with different number of layers have been studied under the conditions of uniaxial pressure. There have been measured current-voltage characteristics at the room temperature in the voltage range of 0 through 1.5 V at 3 magnitudes of the uniaxial pressure from 5 through 25 bars. The current-voltage characteristics are shown to be non-linear and to form a loop during successive increasing and decreasing voltage. The long-term decays of the current on applying voltage to the sample and of the voltage on the sample in the no-load mode on switching off have been observed. The observed results are explained by the model of the surface charge accumulation causing polarization.

**Introduction**

The samples of 2D-MoS$_2$ flakes both mono and multilayered have been intensively investigated because of their great perspectives for nanoelectronics, optoelectronics, biophysics and other applications [1-5]. The electric conductivity in such material was studied in the range of small voltage biases [6-13], and these results may be briefly summarized as follows. First, three kinds of samples were explored. The 1-st include pellets made by pressing the MoS$_2$ flakes in a cylindric or rectangular die, the pressure being 10 through 100 MPa. The 2-nd and 3-rd are in fact single crystalline obtained by different methods (direct chemical vapor deposition or by exfoliation from preliminary obtained ones). It is established that for all the kinds the conductivity may bear different character in dependence of fabrication conditions, doping and modulation charge carriers concentration, defects. In the wide range of low temperatures it usually bears the hopping character both with a constant activation energy and with the variable range hopping mechanism. In some cases the clusters with the quasi-metallic or hopping kind conductivity may appear and aggregate a percolation network determining the total conduction between electric contacts. The uniaxial pressure strongly enhances anisotropy of conductivity. In some papers it is noted that growth of stress applied to a sample leads to a weakening temperature dependence of conductivity. In the case of the pellet samples made of flakes powder by strong pressing the current-voltage characteristics (CVC) were linear while in other kind samples the CVC were both linear and non-linear. For the samples of monolayer flakes the non-linear CVC may be caused by contribution of the metal electric contact forming the Schottky barrier. At the same time the electric conduction in the 2D-MoS$_2$ flake powder samples not subjected to strong pressure, being destructive or changing their structure, while possessing conduction in measurable limits and possible bistability and reversibility of characteristics was not studied yet. This case is of interest for fundamental studies of properties of such materials with a view of possible applications.

The present work is devoted to investigation of the CVC nature in the samples of 2D-MoS$_2$ powder of flakes initially not subjected to strong pressure to form a monolithic substance. While to keep conductivity in the measurable limits the samples were investigated under uniaxial pressure within the moderate limits.

**Experimental details**

The samples for investigation were made of 2D-MoS$_2$ powder of flakes[1] with the average size of a few μm. The samples were shaped as disks, being the 4 mm in diameter and 150 μm thick. The general scheme of measurements is shown in Fig.1. The samples were placed between metallic plungers which were electric contacts and serve to perform uniaxial pressure in the range no more than 2.5 MPa. The Raman spectra measured on the samples under study before and after applying pressure were identical even at maximal pressure evidencing that irreversible change in the sample structure and damages do not appear in the used pressure range.[2] The sample with plungers was place in the cylindric teflon case. The load resistor used to measure the current through the sample is connected in series to the sample outside the teflon case. The digital multimeters Keithley 2000 to measure voltage drops are connected both to the sample and resistor. Scanning of the voltage applied to the sample was made the voltage supply unit PSP-603 with the 20 mV step. Both multimeters and power supply unit were connected to the computer to record results. The measurements of CVC were performed at the room temperature at the 3 fixed pressure magnitudes: 0.5, 1.5 and 2.5 MPa. Also we explored the temporal dynamics of the current through the sample on switching on the applied voltage and decay of the voltage drop on the sample in the no-load mode on switching off the power supply.

---

[1] Produced by Sigma-Aldrich. Product name: Molybdenum(IV) sulfide – powder, <2 μm, 98 %.

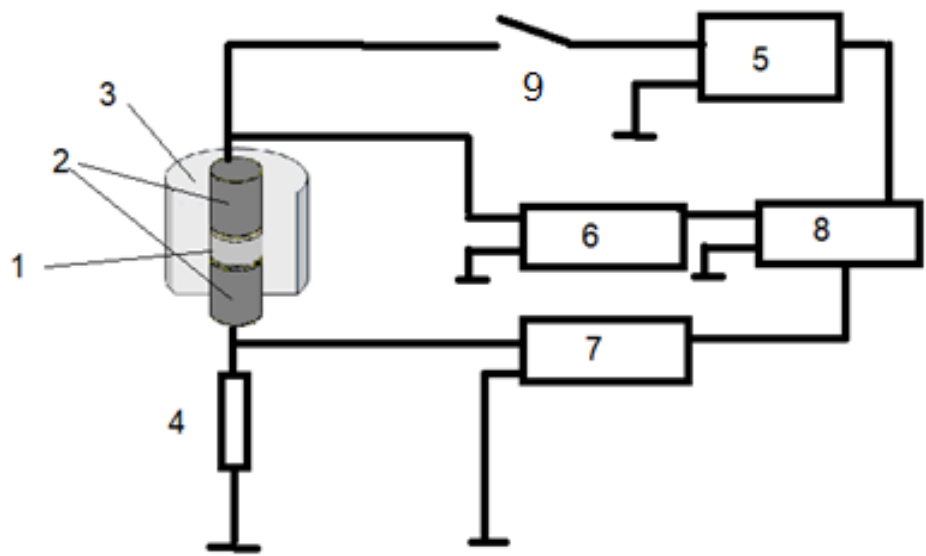

**Fig.1.** The electric scheme to measure the current-voltage characteristics. 1 – sample, 2 – plungers to perform the uniaxial pressure and serve as electric contacts, 3 – cylindric Teflon case, 4 – load resistor to measure the current though the sample, 5 – dc power supply unit, 6, 7 – digital multimeters to measure the voltage drops on the sample and load resistor, 8 – computer, 9 – switch.

**Results and discussion**

Depicted in Fig.2 is a typical CVC of the sample in voltage range of ±1 V which corresponds to the average electric field strength in the sample no more than 67 V/cm. This CVC was measured at the middle value of pressure 1.5 MPa and reflects main common features observed in the whole used pressure range. It is important that measurement of current at each point on the voltage scale was performed after temporal delay about a couple minutes. As is shown below the current during this delay relaxes to a steady value. The arrows near curves show the direction of the applied voltage changing. With growing pressure applied to the sample the CVC curve shifts to a lower voltage. It is seen that in general the CVC curves are non-linear. Besides, the curves corresponding to the increasing and decreasing applied voltage diverge. In the backward branch of lowering voltage the current is larger with the strongest discrepancy of the forward and backward branches being in the middle interval in the voltage scale.

---
[2] The Raman spectra were recorded in the Institute of semiconductor physics, NAS of Ukraine by Dr Nikolenko.

The same CVC curves in the enlarged scale near zero voltage bias applied to the sample (within ±100 mV) are shown in Fig.3. Here one can see a very weak hysteresis in CVC at the zero bias. Initially the voltage drop on the sample increases beginning from zero. However, in the backward CVC branch on the full current decrease down to zero the small residual voltage about 20 mV remains on the sample. On switching to the opposite bias polarity, the CVC curve at a small voltage bias returns to the practically linear run intercepting zero in the current and voltage axes. In the backward branch the CVC again intersects 20 mV in the voltage axis. Such a loop-like behavior of CVC both in the middle voltage range and in the zero vicinity may be considered like evidence of polarization caused by charge redistribution in transverse direction of electric field in the sample. With a view to reveal properties of this process we investigated relaxation of current through the sample on applying voltage to the sample and relaxation the voltage drop on the sample in the no-load mode on disconnecting the sample circuit after long enough duration of the applied voltage.

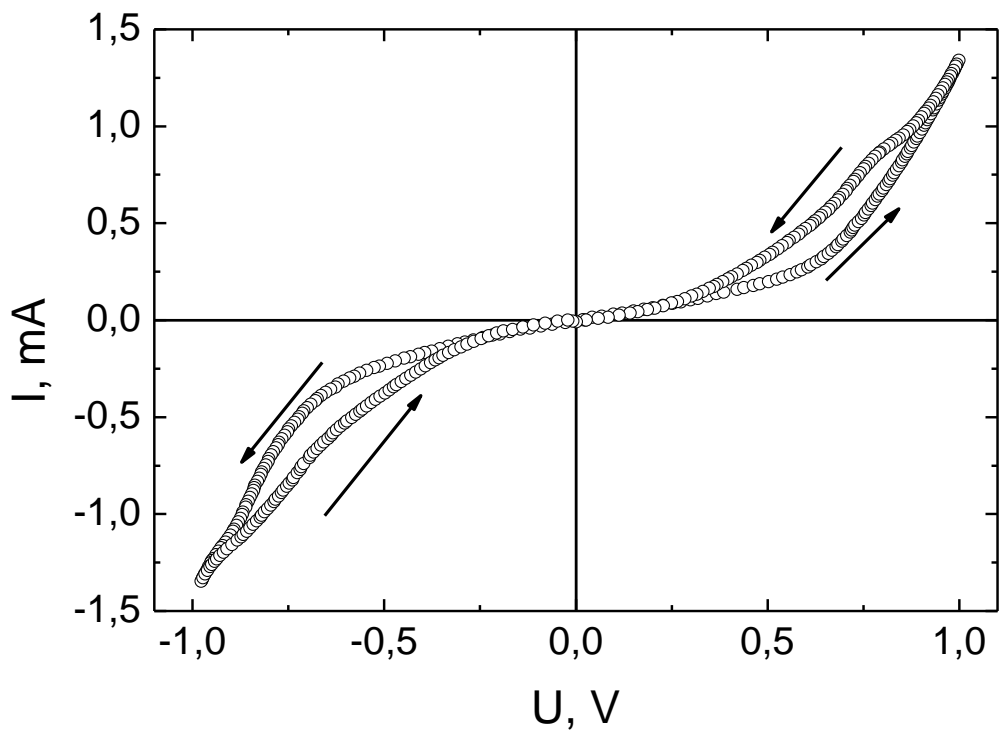

**Fig.2.** The typical CVC of the sample at T=300 K and P=1.5 MPa. The arrows show directions of the applied voltage scanning.

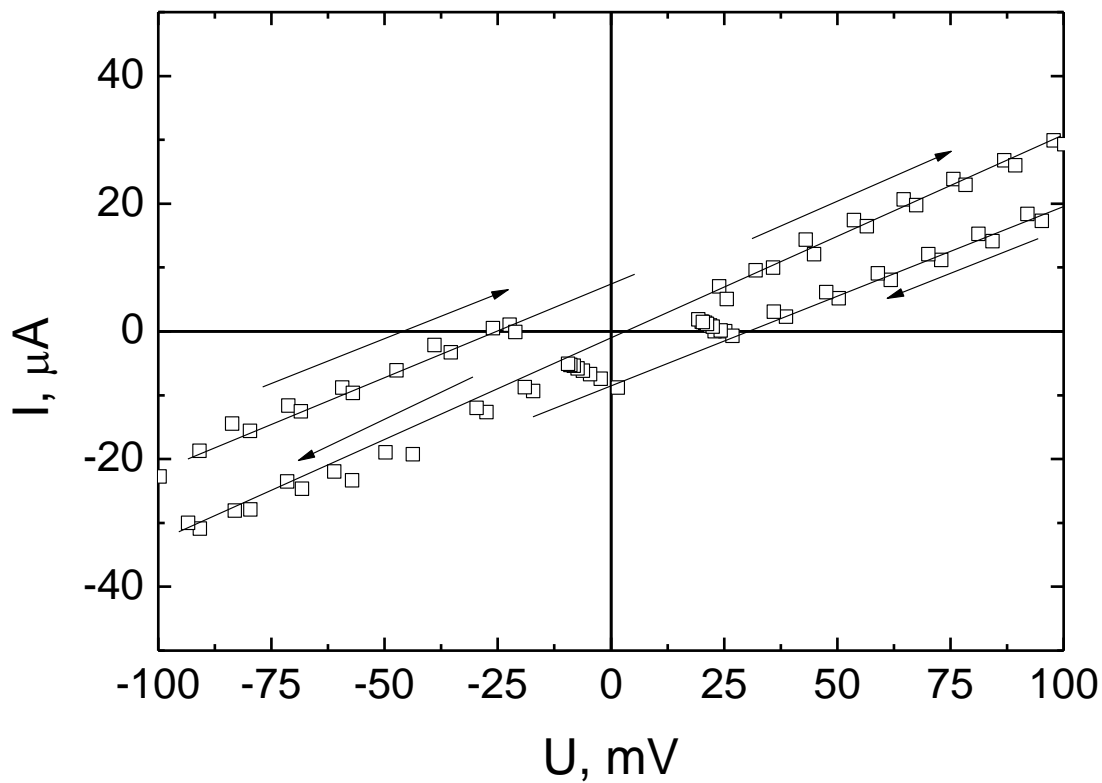

**Fig.3.** The fragment of CVC shown in Fig.2 in the region of small voltage values. The arrows show directions of the applied voltage scanning.

Shown in Fig.4 in the semilogarithmic scale are curves of the current decay in the sample during the constant voltage pulse applied to the sample and of the voltage across the sample decay on disconnecting the sample circuit, the voltage pulse magnitude and duration being 200 mV and 50 s, respectively, at 3 values of pressure: 0,5, 1.5 and 2.5 MPa. It is seen that on disconnecting circuit a long-term decay of voltage across the sample is observed. On the other hand on applying a constant voltage to the sample one observes a long-term decay of the current to some residual magnitude. Such behavior is characteristic for charging and discharging a capacitor. In our case it might be explained, for example, within the model frames proposed and investigated in [14], where it was shown that quite a large concentration of electron states may appear on the surface of the 2D-$MoS_2$ flakes because of the formation of sulfur vacancies during a slow desulfurization process. These states may accumulate a large amount of electrons on the not protected surface. It is obvious that filing and emptying these states depends on band bending by the applied transverse electric field. Let us note that quite long duration of the process makes necessary in measuring CV to keep enough time delay between successive points of voltage. As shown in [14] the surface states are capable to course electron accumulation on the non-protected surface. Besides, as it is seen from our results this effect depends also on the applied pressure. At the stronger pressure relaxation accelerates. It may be caused, for example, by more tight location of flakes or a kind of their reversible deformation which facilitates charge exchange between flakes.

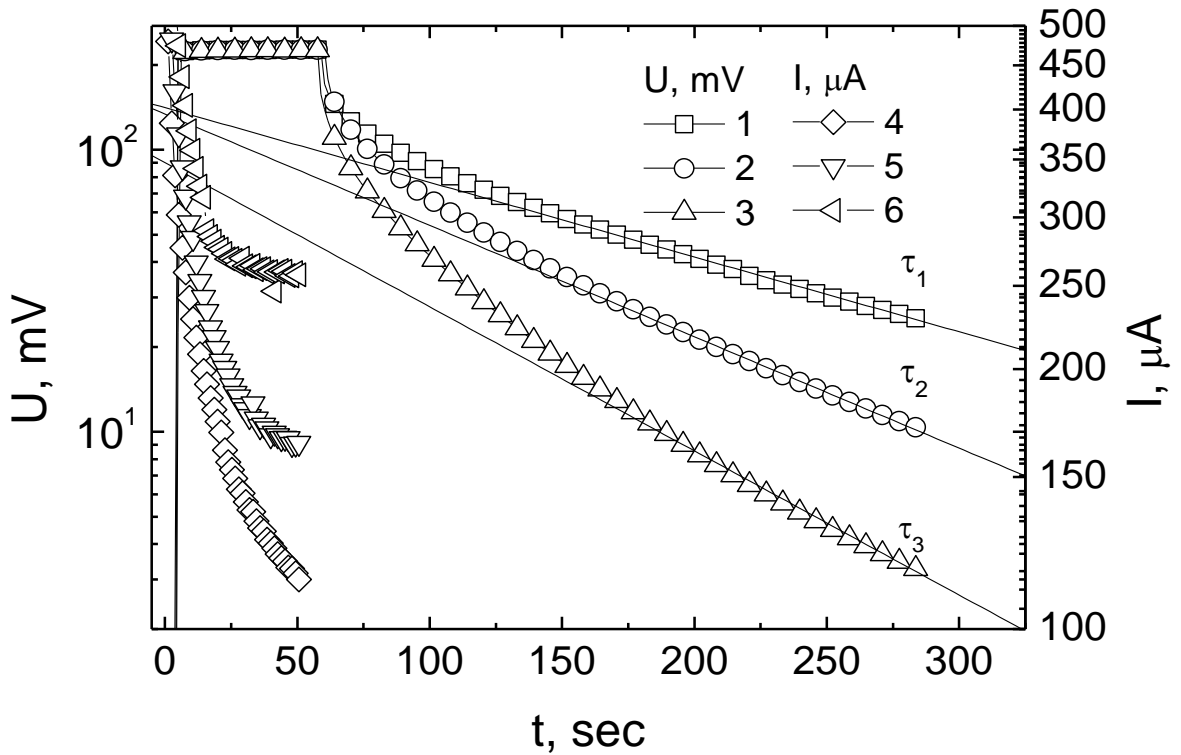

**Fig.4.** The decay curves of the transverse voltage across the sample on disconnecting its circuit in the no-load mode and the current decay during the applied voltage pulse. T=300 K, the voltage pulse magnitude - 200 mV, duration – 50 s, pressure P, MPa: 1,4 – 0.5, 2,5 – 1.5, 3,6 – 2.5. $\tau_1$=164 s, $\tau_2$=114 s, $\tau_3$=85 s.

**Conclusion**

Investigations of CVC of the of samples of the 2D-$MoS_2$ nanoflakes powder under the conditions of the uniaxial pressure in the range of 0.5 through 2.5 MPa have shown that such samples manifest the long-term relaxation of current on applying voltage and a long-term relaxation of voltage across the sample in the no-load mode on disconnecting circuit. The CVCs measured complying a corresponding time delay at each voltage point possess a loop-like shape with different forward and backward voltage run branches. A small residual voltage on the sample is also remained on the full return of current to zero in the backward branch. An increase of the uniaxial pressure leads to relaxation acceleration both current and voltage. The results find explanation within the frames of the model proposed in [14] describing accumulation of the surface charge in the states arising in result of a slow desulfurization process at surface.

**Acknowledgement.** The work was supported by National Research Foundation of Ukraine (Grant application 2020.02/0027). The authors also thank to Dr Morozovska for helpful discussion and Dr Nikolenko for measurement of the Raman spectra.